\documentclass[fleqn,usenatbib]{mnras}
\usepackage{psfrag}

\usepackage{etoolbox}
\makeatletter
\newcount\c@additionalboxlevel
\newcount\c@maxboxlevel
\patchcmd\@combinedblfloats{\box\@outputbox}{%
  \stepcounter{additionalboxlevel}%
  \box\@outputbox
}{}{\errmessage{\noexpand\@combinedblfloats could not be patched}}

\AtBeginShipout{%
  \ifnum\value{additionalboxlevel}>\value{maxboxlevel}%
    \typeout{Warning: maxboxlevel might be too small, increase to %
      \the\value{additionalboxlevel}%
    }%
  \fi 
  \@whilenum\value{additionalboxlevel}<\value{maxboxlevel}\do{%
    \typeout{* Additional boxing of page `\thepage'}%
    \setbox\AtBeginShipoutBox=\hbox{\copy\AtBeginShipoutBox}%
    \stepcounter{additionalboxlevel}%
  }%
  \setcounter{additionalboxlevel}{0}%
}
\makeatother

\usepackage{mathptmx}

\usepackage[T1]{fontenc}
\usepackage{ae,aecompl}

\usepackage{upgreek}
\usepackage{amsmath}
\usepackage{amssymb}
\usepackage{xspace}
\usepackage{rotating}
\usepackage{pdflscape}
\usepackage{threeparttable}

\title[Star formation in cluster dwarf galaxies]{Star formation in low-redshift cluster dwarf galaxies}

\author[C. M. Rude et al.]{
Cody M. Rude,$^{1}$
Madina R. Sultanova,$^{1}$
Gihan L. Ipita Kaduwa Gamage,$^{1}$
\newauthor
Wayne A. Barkhouse$^{1}$\thanks{E-mail: wayne.barkhouse@und.edu}
and Sandanuwan P. Kalawila Vithanage$^{2}$
\\
$^{1}$Department of Physics and Astrophysics, University of North Dakota, Grand Forks, ND 58202, USA\\
$^{2}$Department of Physics, University of Ruhuna, Matara, Sri Lanka 
}

\date{Accepted XXX. Received YYY; in original form ZZZ}

\pubyear{2020}

\begin{document}
\label{firstpage}
\pagerange{\pageref{firstpage}--\pageref{lastpage}}
\maketitle

\begin{abstract}
Evolution of galaxies in dense environments can be affected by close encounters with neighbouring galaxies and 
interactions with the intracluster medium. Dwarf galaxies (dGs) are important as their low mass makes them more susceptible 
to these effects than giant systems. Combined luminosity functions (LFs) in the $r$- and $u$-band of 15 galaxy 
clusters were constructed using archival data from the Canada-France-Hawaii Telescope. LFs were measured as a 
function of cluster-centric radius from stacked cluster data. Marginal evidence was found for an increase in the faint-end 
slope of the $u$-band LF relative to the $r$-band with increasing cluster-centric radius. The dwarf-to-giant ratio (DGR) was found 
to increase toward the cluster outskirts, with the $u$-band DGR increasing faster with cluster-centric radius compared to the $r$-band. 
The dG blue fraction was found to be $\sim 2$ times larger than the giant galaxy blue fraction over all cluster-centric 
distance ($\sim 5\sigma$ level). The central concentration ($C$) was used as a proxy to distinguish nucleated versus non-nucleated dGs. 
The ratio of high-$C$ to low-$C$ dGs was found to be $\sim 2$ times greater in the inner cluster region compared to the outskirts ($2.8\sigma$ level). 
The faint-end slope of the $r$-band LF for the cluster outskirts ($0.6 \leq r/r_{200} < 1.0$) is steeper than the SDSS 
field LF, while the $u$-band LF is marginally steeper at the $2.5\sigma$ level. Decrease in the 
faint-end slope of the $r$- and $u$-band cluster LFs towards the cluster centre is consistent with quenching of star 
formation via ram pressure stripping and galaxy-galaxy interactions. 
\end{abstract}



\section{Introduction}

Galaxies reside in a wide range of environments, from voids to the centre of massive clusters. Galaxies in clusters 
differ both  morphologically \citep[e.g.][]{Dressler1980,Dressler1997} and in their star formation history compared with 
field galaxies \citep[e.g.][]{Quadri2012,Darvish2018}. The majority of galaxies in present day clusters are 
passively evolving, with approximately 20 per cent showing evidence of some star formation \citep[e.g.][]{Poggianti06}. 
Based on a sample of 32 low-redshift galaxy clusters ($0.04< z <0.07$) selected from the WINGS \citep{Cava09} and OmegaWINGS surveys 
\citep{Gullieuszik15}, \citet{Paccagnella2017} found that $7.2\pm 0.2\%$ of $V< 20$ cluster galaxies within 1.2 virial radii 
appear to be post-starburst \citep[see also][]{Paccagnella2019}. 

A typical galaxy cluster contains a passive population of elliptical/S0 galaxies that form a red-sequence ridge-line 
in the colour-magnitude plane \citep[e.g.][]{Lopez2004}. Blue cloud galaxies occupy the region blueward of the 
red-sequence, while the colour space between the red-sequence and the blue cloud contains green valley galaxies 
\citep{Wyder07}. These galaxies may be in the process of being quenched and moving from the blue cloud 
into the red-sequence region \citep{Bremer2018,Belfiore2018}. Conversely, some red-sequence galaxies may have 
recently had their star formation reignited and are transitioning back to the blue cloud \citep{Coenda18,Darvish2018}. 

Several mechanisms have been proposed to account for differences between field and cluster galaxies. Ram pressure 
from the intracluster medium (ICM) acting on galaxies as they move through a cluster, is expected to compress or 
even completely remove the ISM from individual galaxies \citep{GunnGott72,Quilis2000,Tonnesen2007}. Alternatively, 
interactions between large neighbouring galaxies \citep[galaxy harassment;][]{Moore1996,Moore1998} could 
morphologically transform disk galaxies into spheroidals. Galaxy starvation or strangulation \citep{Larson1980}, 
where inflowing star-forming gas is truncated, will quench star formation once the gas supply is 
exhausted \citep[e.g.][]{Boselli06,Maier19}. 

Dwarf galaxies are particularly important for studying environmental effects as their low mass makes them more 
susceptible to external influences than massive galaxies. Morphologically, \citet{Sandage84} classified dwarf 
ellipticals as having a flatter surface brightness profile than giant elliptical galaxies. More recently, S0 galaxies 
have been classified in parallel with spiral galaxies \citep{Cappellari2011,Kormendy2012}. \citet{Cappellari2011} 
place dwarf ellipticals (which they refer to as dwarf spheroidals) at the end of the S0 sequence, which mirror dwarf 
irregular galaxies at the end of the spiral sequence. 

In the local universe, there are signs of star formation in early-type cluster dwarf galaxies 
\citep[e.g.][]{Caldwell1998,Lisker06,DeRijcke10,Urich17}. For example, \citet{Hamraz19} used HST ACS photometry of Virgo, Fornax, 
and Coma clusters to estimate the fraction of early-type dwarf galaxies containing young stellar populations; $15\pm 3\%$ for the 
Virgo cluster, $11\pm 2\%$ for Fornax, and $2\pm 1\%$ for the Coma cluster. 

\citet{Barkhouse07} combined 
the luminosity functions (LFs) of 57 low-redshift clusters and found that the faint-end of the LF is sensitive to distance 
from the cluster centre. The slope of the faint-end is important as it contains information about the relative number of dwarf 
galaxies in the cluster. \citet{Beijersbergen2002} measured deep LFs for various cluster-centric regions of the Coma cluster. In the 
outer region of the cluster they found that the faint-end of the $u$-band LF has a steeper slope relative to the $r$-band. 
One possible explanation for the difference in slope in the cluster outskirts is the presence of a population of 
star-forming dwarf galaxies.

The spectroscopic measurement of the LF of A85 from \citet{Agulli2016} showed no statistically significant steepening 
of the faint-end slope for red galaxies from the core to the outskirts of the cluster. The contribution of the faint 
red galaxy component was found to dominate more at the outer regions of the cluster compared to the central area. 
The blue, star forming galaxies were found to make up a small fraction of the cluster population at small cluster-centric 
distances, but are roughly equal in number to red galaxies at the cluster outskirts. For A2151, a spiral-rich cluster, 
\citet{Agulli2017} found that the faint-end of the spectroscopic LF is independent of cluster-centric distance from the core 
to the outskirts of the cluster. This study also found a deficit of red dwarf galaxies.

We present Canada-France-Hawaii Telescope (CFHT) $u$- and $r$-band measurements of LFs, blue fractions,
dwarf-to-giant ratios (DGRs), and a comparison of the central concentration of individual dwarf galaxies for 15 Abell 
clusters as a function of cluster-centric radius. Observations and data reduction are described in 
Section \ref{sec:data_reduction}, and LFs are presented in Section \ref{sec:luminosity_functions}. 
DGRs and blue fractions are given in Section \ref{dwarf-to-giant_ratios_and_blue_fraction}. 
In Section \ref{morphology} we describe the morphological properties of dwarf galaxies using the central concentration 
statistic. Our discussion and conclusions are given in Section \ref{discussion}. 

The cosmological parameters of $\mathrm{H_0}=70~\mbox{km}~\mbox{s}^{-1}~\mbox{Mpc}^{-1}$, $\Omega_{\Lambda}=0.7$, and 
$\Omega_{M}=0.3$ are used throughout this paper.

\section{Data Reduction}
\label{sec:data_reduction}
Data for this study consists of archival observations from the 3.6 metre CFHT imaged with the MegaPrime/MegaCam CCD 
camera. All clusters have $u$- and $r$-band data available with adequate exposure times to allow sampling of 
the dwarf galaxy population for our chosen clusters. The adapted redshift range ($0.03<z<0.184$) ensures that cluster 
dwarf galaxies are spatially sampled out to the virial radius within the one square degree field-of-view of the 
telescope+detector. A summary of cluster observations is given in Table \ref{target_overview}, while a detailed 
overview of the data reduction process can be found in \citet{Rude2015}.

\begin{table*}
\caption{Abell cluster sample.}
\centering
\begin{threeparttable}
\centering
\begin{tabular}{| c | c | c | c | c | c | c | c | c | c | c | c |}
\hline
Cluster & RA\tnote{a}  & Dec\tnote{a} & $z$ & Exposure ($r$) & Exposure ($u$) & Mag Limit ($r$) & Mag Limit ($u$)\\
& (deg) & (deg) & & (s) & (s) & & \\ \hline	 
A76   &  9.9832 &	 6.8486 &	0.041 &	 240 &  1200 & 22.7 & 23.9\\
A98N  &	 11.6031 &	20.6218 &	0.104 &	2160 &	2160 & 22.5 & 23.8\\
A98S  &	 11.6221 &	20.4680 &	0.104 &	2160 &	2160 & 22.5 & 23.8\\
A350  &	 36.2721 &	-9.8366 &	0.159 &	2000 &	3000 & 24.3 & 25.2\\
A351  &	 36.3331 &	-8.7218 &	0.111 &	2000 &	4200 & 24.3 & 25.2\\
A362  &	 37.9215 &	-4.8827 &	0.184 &	2500 &	3000 & 24.3 & 25.2\\
A655  &	126.3712 &	47.1337 &	0.127 &	2940 &	3320 & 23.8 & 24.7\\
A795  &	141.0222 &	14.1727 &	0.136 &	2880 &	 700 & 23.0 & 24.2\\
A1920 &	216.8524 &	55.7502 &	0.131 &	4000 &	6000 & 24.3 & 25.2\\
A1940 &	218.8686 &	55.1312 &	0.140 &	2000 &	3000 & 24.2 & 25.1\\
A2100 &	234.0773 &	37.6438 &	0.153 &	1600 &	1600 & 24.2 & 24.9\\
A2107 &	234.9127 &	21.7827 &	0.041 &	 600 &	3600 & 23.0 & 24.1\\
A2147 &	240.5709 &	15.9747 &	0.035 &	 600 &	3060 & 23.2 & 24.3\\
A2199 &	247.1594 &	39.5513 &	0.030 &	1600 &	1600 & 23.9 & 25.0\\ 
A2688 &   0.0318 &      15.8342 &       0.151 & 2160 &  2160 & 22.4 & 23.7\\ \hline
\end{tabular}
\begin{tablenotes}
  \item[a] J2000.0
\end{tablenotes}
\end{threeparttable}
\label{target_overview}
\end{table*}

Data from the CFHT archive were pre-processed via bias-subtraction and flat-fielding. Total exposure times listed 
in Table \ref{target_overview} are the sum of integration times of individual images. To create a final 
calibrated image, individual exposures were median combined using the software packages \textsc{Source Extractor} 
\citep{Bertin96}, \textsc{Scamp} \citep{Bertin06}, and \textsc{SWarp} \citep{Bertin02}. Once final images were 
produced, the \textsc{Picture Processing Package} \citep[\textsc{PPP};][]{Yee1991} was used to create an object 
catalogue with measured magnitudes for each cluster image. Object classification was performed using \textsc{PPP's} 
built-in object classifier ($C_{2}$). Each object was classified by comparing the flux ratio of its inner and outer 
regions with that of reference stars \citep{Yee1991}. 

To minimise filter-selection bias, object detection was performed independently in both the $r$- and $u$-band 
\citep[for a similar procedure see][]{Beijersbergen2002}, 
and the resultant catalogues visually inspected for bogus detections and missed objects. Since a main goal 
of this study is to look for the effect on star formation as dwarf galaxies fall into the cluster environment, 
it is important to construct $u$- and $r$-band LFs so that dwarf galaxies detected in the $u$-band but 
not in the $r$-band are included in the $u$-band LF. Given that for each cluster our $u$-band images are $\sim 1$ mag deeper 
than our $r$-band data (see Table \ref{target_overview}), our object detection procedure will not have an adverse impact 
on the construction of our cluster LFs.

Once object detection and cleaning was complete, catalogues were merged together to form a master catalogue. 
The $r$-band zero points were calibrated using SDSS {\tt PSF} magnitudes of stars. For the $u$-band, this method resulted 
in a magnitude offset between SDSS galaxies and our sample. Therefore, the $u$-band was calibrated by comparing {\tt CMODEL} 
galaxy magnitudes. The magnitude limit for each image was chosen to be 0.8 mag brighter than the turnover in galaxy counts 
versus apparent magnitude, or when $\sigma_{u-r}=0.2$ mag (whichever is brighter), to ensure 100 per cent completeness (see 
Table \ref{target_overview}).

\section{Luminosity Functions}
\label{sec:luminosity_functions}

LFs were constructed for each cluster by first fitting a straight line to the cluster red-sequence using linear least squares. 
This fit was generally carried out using galaxies within a radius of 1 Mpc from the cluster centre. In cases where the 
red-sequence was difficult to discern, a smaller radial cut was used. 

A background-corrected rectified colour histogram was computed for each galaxy cluster. The colour of each galaxy 
was offset to remove the slope of the red-sequence. The dispersion of the red-sequence was determined by fitting a 
Gaussian function to the colour histogram. The background of each cluster was measured from an area $>3$ Mpc from 
the cluster centre using the outskirts of the cluster image. For the lowest redshift clusters, a composite background 
was made from the backgrounds of the remaining clusters in our sample. 

Once the red-sequence was measured, the $\lambda$ richness parameter \citep{Rykoff2012} was calculated using the 
$u-r$ colour and $r$-band magnitude of each galaxy. This measurement requires an estimate of $m^*_r$, which was 
determined using $M_r^*=-21.47$ (from Barkhouse et al. (2007) converted to the SDSS $r$-band), and an evolution and 
k-correction, which was calculated using \textsc{GALAXEV} \citep{Bruzual03}. Following \citet{Rykoff2012}, the 
$\lambda$ richness measurement was computed using a counting radius of 0.9 Mpc, with a correction applied for chip 
gaps. 

\begin{figure}
\includegraphics[width=\linewidth]{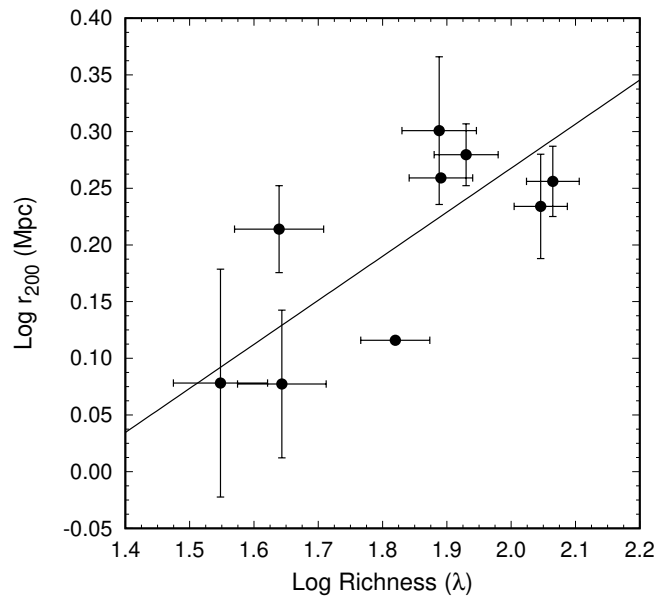}
\caption{Logarithmic correlation between $r_{200}$ and $\lambda$, where $r_{200}$ was determined from velocity dispersion measurements.}   
\label{lambda_r200}
\end{figure}

For clusters with a measured velocity dispersion ($\sigma_{v}$), $r_{200}$ was determined from 
$r_{200}=\sqrt{3}\sigma_{v}/10 H(z)$ \citep{Carlberg1997}. Using the BCES method of \citet{Akritas96}, a linear fit 
between $\lambda$ and $r_{200}$ in logarithmic space yields ${\rm log}\,r_{200}=(0.39\pm 0.10)\,{\rm log} \lambda - (0.51\pm 0.19)$. 
This relationship was used to estimate $r_{200}$ for those clusters without a published $\sigma_{v}$ (see Fig.~\ref{lambda_r200}). 
The values for $\lambda$ and $r_{200}$ for the cluster sample are tabulated in Table \ref{data_overview}. 

\begin{table*}
\caption{Measured cluster properties. Column 1 gives the cluster name, Column 2 is the red-sequence 
dispersion, Column 3 gives the slope of the red-sequence fit, Column 4 is the y-intercept of the red-sequence fit, 
Column 5 is the richness parameter, Column 6 is the velocity dispersion, Column 7 is the velocity dispersion source, and 
Column 8 gives $r_{200}$.}
\centering
\begin{threeparttable}
\centering
\begin{tabular}{ | c | c | c | c | c | c | c | c |}
\hline
\rule{0pt}{2.5ex} Cluster & $\sigma_{RS}$ & Slope & Y-Intercept & $\lambda$ & $\sigma_v$ & $\sigma_v$ Reference & $r_{200}$ \\
\rule{0pt}{2.5ex} & & & & & ($\mbox{km}~\mbox{s}^{-1}$) & & (Mpc) \\ \hline
\rule{0pt}{2.5ex}   A76 & 0.074 & $-0.132 \pm 0.008$ & $4.56 \pm 0.12$ & $44.0\pm 7.0 $ & $492\pm 145$ &  Huchra et al. (2010) &$1.19 \pm 0.18$ \\ 
\rule{0pt}{2.5ex}  A98N & 0.093 & $-0.065 \pm 0.018$ & $3.45 \pm 0.32$ & $57.3\pm 8.9 $ & & & $1.49 \pm 0.92$\tnote{a} \\ 
\rule{0pt}{2.5ex}  A98S & 0.089 & $-0.103 \pm 0.021$ & $4.21 \pm 0.37$ & $116.8\pm 12.3$ & & & $1.97 \pm 1.31$\tnote{a} \\ 
\rule{0pt}{2.5ex}  A350 & 0.033 & $-0.100 \pm 0.014$ & $4.57 \pm 0.28$ & $26.5\pm 5.2 $ & & & $1.10 \pm 0.63$\tnote{a} \\ 
\rule{0pt}{2.5ex}  A351 & 0.080 & $-0.111 \pm 0.016$ & $4.68 \pm 0.30$ & $35.3\pm 6.0 $ & $510\pm 118$ & Popesso et al. (2007) & $1.20\pm 0.28$ \\ 
\rule{0pt}{2.5ex}  A362 & 0.102 & $-0.093 \pm 0.012$ & $4.61 \pm 0.24$ & $90.4\pm 9.6 $ & & & $1.78 \pm 1.16$\tnote{a} \\ 
\rule{0pt}{2.5ex}  A655 & 0.084 & $-0.105 \pm 0.012$ & $4.44 \pm 0.23$ & $111.2\pm 10.6$ & $736\pm 78$ & Popesso et al. (2007) & $1.71\pm 0.18$ \\ 
\rule{0pt}{2.5ex}  A795 & 0.083 & $-0.089 \pm 0.010$ & $4.28 \pm 0.19$ & $116.1\pm 11.0$ & $778^{+61}_{-50}$ & Rines et al. (2013) & $1.80 \pm 0.13$ \\ 
\rule{0pt}{2.5ex} A1920 & 0.078 & $-0.103 \pm 0.012$ & $4.55 \pm 0.22$ & $66.0\pm 8.2 $ & $562$ & Tovmassian \& Andernach (2012) & $1.31$ \\ 
\rule{0pt}{2.5ex} A1940 & 0.070 & $-0.110 \pm 0.010$ & $4.67 \pm 0.19$ & $77.8\pm 8.8 $ & $785$ & Struble \& Rood (1999) & $1.82$ \\ 
\rule{0pt}{2.5ex} A2100 & 0.072 & $-0.095 \pm 0.009$ & $4.42 \pm 0.17$ & $54.6\pm 7.6 $ &  & & $1.46\pm 0.90$\tnote{a} \\ 
\rule{0pt}{2.5ex} A2107 & 0.051 & $-0.103 \pm 0.009$ & $4.15 \pm 0.14$ & $43.6\pm 7.0 $ & $674^{+67}_{-52}$ & Oegerle \& Hill (2001) &$1.64\pm 0.14$ \\ 
\rule{0pt}{2.5ex} A2147 & 0.077 & $-0.120 \pm 0.007$ & $4.33 \pm 0.12$ & $77.3\pm 10.3$ & $821^{+68}_{-55}$ & Barmby \& Huchra (1998) & $2.00\pm 0.30$ \\ 
\rule{0pt}{2.5ex} A2199 & 0.105 & $-0.124 \pm 0.005$ & $4.47 \pm 0.08$ & $85.1\pm 9.7 $ & $780^{+52}_{-44}$ & Oegerle \& Hill (2001) & $1.90\pm 0.12$ \\ 
\rule{0pt}{2.5ex} A2688 & 0.106 & $-0.090 \pm 0.010$ & $4.39 \pm 0.20$ & $34.1\pm 5.9 $ &  & & $1.22 \pm 0.71$\tnote{a} \\ \hline
\end{tabular}
\begin{tablenotes}
  \item[a] $r_{200}$ is calculated using the relationship between $r_{200}$ and $\lambda$ given in Section \ref{sec:luminosity_functions}.
\end{tablenotes}
\end{threeparttable}
\label{data_overview}
\end{table*}
 
\begin{figure*}
\includegraphics[width=\linewidth]{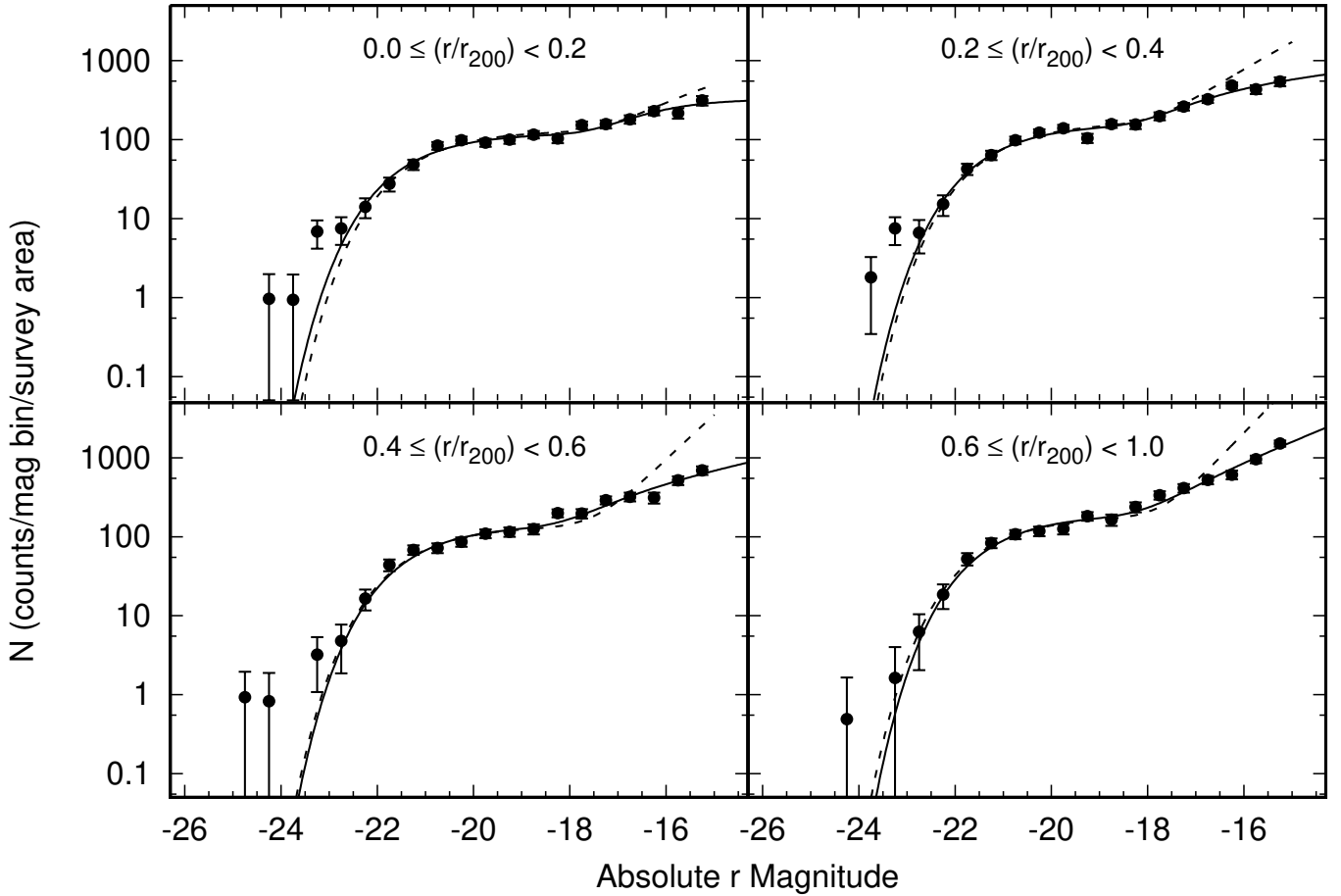}
\caption{The $r$-band LFs from four cluster-centric regions for the combined sample of 15 Abell clusters. The solid lines 
are the fit, and the dashed lines are from \citet{Barkhouse07}. The results from Barkhouse et al. have been scaled to 
match the bright-end of our LFs.}
\label{rband_lfs}
\end{figure*}

A k-correction was applied to each galaxy based on $u-r$ colour following \citet{Chilingarian10} and 
\citet{Chilingarian12}. All galaxies $>3\sigma$ redward of the red-sequence were excluded. The presence 
of distant background clusters in the field-of-view of our target clusters will tend to skew the LF, 
especially at the faint-end where the background cluster galaxy counts will add directly to the LF. 
To minimise this effect, we exclude galaxies that have a colour $>3\sigma_{u-r}$ redward of the red-sequence 
for each cluster. Due to the low redshift of our cluster sample, we expect foreground contamination from 
lower redshift clusters to be minimal. The adopted colour selections were applied equally to both the cluster and 
associated background fields.

For the composite LF, individual LFs for each cluster were measured and then combined following the procedure in \citet{Schechter1976}. 
The uncertainty for the $i$th magnitude bin of the background-corrected LF was calculated using 
$\sqrt{N_{ci} + 1.69 \ N_{bi}}$, where $N_{ci}$ is the number of background-subtracted cluster galaxies, and $N_{bi}$ 
is the number of expected background galaxies in the cluster. The factor of 1.69 is used to account for the 30 per cent 
field-to-field variation in the background counts \citep{Barkhouse07}. 

The cluster sample was divided into four annuli based on $r_{200}$: $0.0 \leq r/r_{200} < 0.2$, $ 0.2 \leq r/r_{200} < 0.4 $, 
$0.4 \leq r/r_{200} < 0.6$, and $0.6 \leq r/r_{200} < 1.0$. The LFs were fit using a double Schechter function following 
the procedure outlined in \cite{Barkhouse07}. The double Schechter function is given by

\begin{equation}
\begin{split}
N(M) dM = k \Big( & N^*_1 e^{k(\alpha_1 + 1)(M^*_1-M) - \exp[k(M^*_1-M)]} + \\ 
& 2N^*_1 e^{k(\alpha_2 + 1)(M^*_2-M) - \exp[k(M^*_2-M)]}\Big)dM,
\end{split}
\end{equation}
where $k=0.4 \ln 10$. Since there are degeneracies when fitting a double Schechter function, $M^*_1$ was determined 
by fitting a single Schechter function to the bright-end with the faint-end slope fixed at $\alpha_{1}=-1$. 

Non-linear least squares fit results for the $r$-band LFs are shown in Fig.~\ref{rband_lfs} (solid lines), while the 
fit parameters are given in Table \ref{tab:lf_fits}. We find a trend that $M_2^*$ gets 
brighter by 0.98 mag with increasing cluster-centric radius (significant at the $5.2\sigma$ level). In addition, the 
faint-end slope is 1.7 times steeper in the $0.6 \leq r/r_{200} < 1.0$ annulus compared to the $0.0 \leq r/r_{200} < 0.2$ 
region, significant at the $6.7\sigma$ level. This may indicate that the dwarf galaxy population is being disrupted in 
the inner cluster region. 

For comparison, the Schechter function fits from \citet{Barkhouse07} are shown in Fig.~\ref{rband_lfs} (dashed lines). 
The LF parameters were converted to our adopted cosmology, and a correction was applied to convert Cousins $R_c$ 
to Sloan $r$ \citep{Fukugita1995}. The fits from Barkhouse et al. were normalised to the CFHT data by fitting 
a scale factor to the bright-end of the LF. 

The faint-end LF slopes from Barkhouse et al. are steeper than our values. This discrepancy may be due to 
differences in cluster samples and background corrections. The sample median redshift for Barkhouse et al. is $z=0.06$, 
while the median redshift for our study is $z=0.13$. The two cluster samples are similar in terms of richness 
when comparing $r_{200}$ values using the same distance scale.

\begin{table*}
\caption{Parameters derived from fitting double Schechter functions to the $r$-band LFs.}
\centering
\begin{threeparttable}
\centering
\begin{tabular}{|c|c|c|c|c|c|c|}
\hline
Radial Bin & $M_1^*$  & $\chi^{2}_{\nu}$ & $M_2^*$ & $\alpha_{2}$ & $\chi^{2}_{\nu}$ & No. of Clusters\\ \hline
$0.0 \leq r/r_{200} < 0.2$ & $-21.48 \pm 0.11$ & 0.80 & $-16.47 \pm 0.17$ & $-0.95 \pm 0.10$ & 1.39 & 15 \\ 
$0.2 \leq r/r_{200} < 0.4$ & $-21.41 \pm 0.09$ & 1.35 & $-17.22 \pm 0.13$ & $-1.21 \pm 0.04$ & 2.05 & 15 \\ 
$0.4 \leq r/r_{200} < 0.6$ & $-21.37 \pm 0.10$ & 3.41 & $-17.34 \pm 0.12$ & $-1.37 \pm 0.03$ & 3.58 & 15 \\
$0.6 \leq r/r_{200} < 1.0$ & $-21.33 \pm 0.08$ & 2.17 & $-17.45 \pm 0.08$ & $-1.63 \pm 0.02$ & 3.74 & 13 \\
$0.0 \leq r/r_{200} < 1.0$ & $-21.40 \pm 0.05$ & 3.73 & $-17.36 \pm 0.06$ & $-1.38 \pm 0.02$ & 3.37 & 13 \\ \hline
\end{tabular}
\end{threeparttable}
\label{tab:lf_fits}
\end{table*}

\begin{figure*}
\includegraphics[width=\linewidth]{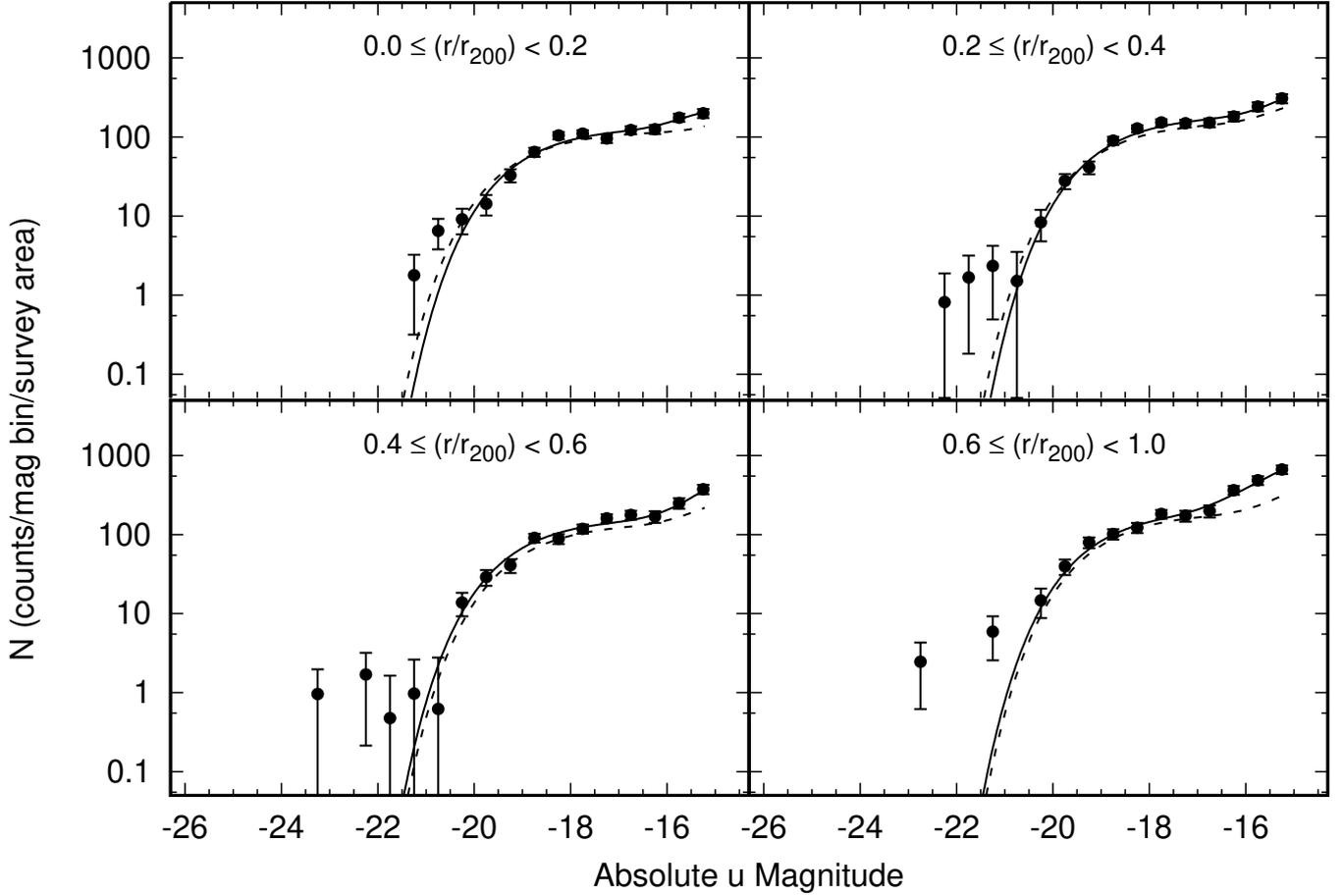}
\caption[The $u$-band LF for four cluster-centric regions.]{The $u$-band LFs from four cluster-centric regions. The 
solid lines are the fit to the $u$-band data, while the dashed lines are the fit to the $r$-band data shifted to fainter 
absolute magnitude by 2.26 mag.}
\label{uband_lfs}
\end{figure*}

The $u$-band LFs for the four cluster-centric radial bins are shown in Fig.~\ref{uband_lfs} (solid lines), and the results 
of the LF fits are given in Table \ref{tab:lf_fits_uband}. The parameters of the second Schechter function are not well 
constrained. Due to the relatively bright cutoff of the $u$-band LF used in fitting the parameters, the value of $N_2^*$ 
is held fixed as the faint-end slope $\alpha$ is the primary parameter of interest. Barkhouse et al. found that the 
geometric mean of $N_2^*/N_1^*=2.12$ for a sample of 57 Abell clusters, thus the value of $N_2^*$ was fixed at $2N_1^*$. 
The integral of the resulting fit is required to match the total number of galaxies in the magnitude interval being measured. 

The $r$-band double Schechter function fits (dashed lines) are compared to the $u$-band after applying a 2.26 magnitude 
shift (Fig.~\ref{uband_lfs}), which is the typical $u-r$ colour of a red-sequence galaxy from our sample at 
$M_{r}=-19.5$. There is a weak trend in which the faint-end of the $u$-band LF becomes steeper than the $r$-band with 
increasing cluster-centric radius. Large uncertainties in the fits of the $u$-band faint-end slope prevent us from making 
any definitive statement regarding the physical significance of such an effect. However, we note that \cite{Beijersbergen2002} 
found that the $u$-band faint-end slope is steeper than the $r$-band in the outskirts of the Coma cluster. Beijersbergen et al. 
attributed this result as possibly due to an enhancement of star formation. This would cause faint dwarf galaxies to become 
brighter in the $u$-band, but be relatively unchanged in the $r$-band, thus yielding an increase in the faint-end slope of 
the $u$-band LF relative to the $r$-band.

\begin{figure*}
\includegraphics[width=\linewidth]{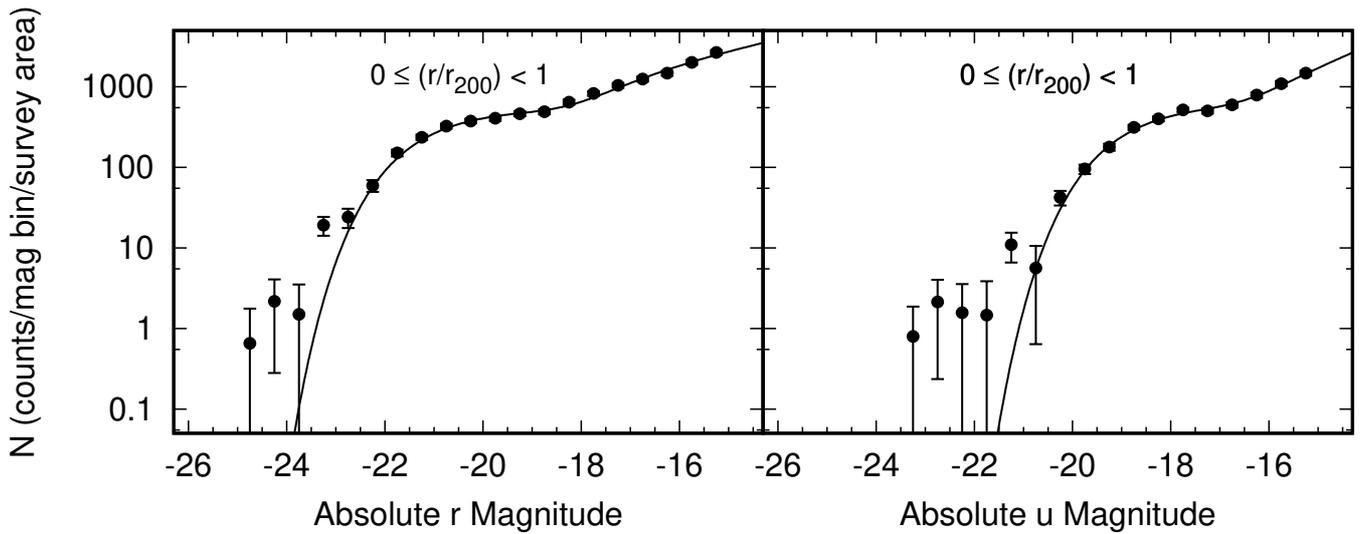}
\caption{The $r$-band (left panel) and $u$-band (right panel) composite LF for $0.0\leq r/r_{200}<1$. A double Schechter 
function fit to each LF is depicted by the solid line. The LF fit parameters are tabulated in Table \ref{tab:lf_fits} for the $r$-band and 
Table \ref{tab:lf_fits_uband} for the $u$-band.}
\label{full_lfs}
\end{figure*}

In addition to constructing LFs for the four radial annuli, we also measured the $r$- and $u$-band LF for a cluster-centric 
distance interval of $0.0\leq r/r_{200}<1$. These ``full'' LFs are depicted in Fig.~\ref{full_lfs} (solid lines) and the fit 
parameters are tabulated in Table \ref{tab:lf_fits} for the $r$-band and Table \ref{tab:lf_fits_uband} for the $u$-band. 
We find that the faint-end slope of the full $u$-band LF is slightly steeper than the $r$-band LF (different at the $1.6\sigma$ level).

Comparing our LFs with other studies can be problematic given differences in cluster sample, background correction, filter bandpass, 
cluster-centric radius coverage, magnitude depth, photometric vs. spectroscopic LFs, etc. To minimise these effects we compared our LF for A2199 with the 
spectroscopic LF of A2199 from \citet{Rines2008} by selecting galaxies using the same criteria; cluster-centric radius $\leq 1.11$ Mpc, 
$M_{r}\leq -15.6$, and excluding the BCG in the LF fit. Fitting a single Schechter function to the $r$-band LF of A2199, we find 
$\alpha=-1.12\pm 0.06$ and $M^{*}=-21.77\pm 0.45$. In comparison, \citet{Rines2008} found $\alpha=-1.13^{+0.07}_{-0.06}$ and 
$M^{*}=-21.11^{+0.21}_{-0.25}$. Due to the agreement between our photometric LF and the spectroscopic LF of A2199, we have confidence 
that our background correction technique used for constructing LFs is valid.  

We compare our full $r$- and $u$-band LFs ($0.0\leq r/r_{200}<1$) with the spectroscopic LF of A85 ($0.0\leq r/r_{200}<1.5$, using our distance scale) from 
\citet{Agulli2016}. The faint-end slope of the LF for red galaxies from Agulli et al. yields $\alpha=-1.53\pm 0.02$, while our $r$-band LF 
has $\alpha=-1.38\pm 0.02$. The difference may be due to sampling out to a greater cluster-centric radius in the Agulli et al. study, where red dwarf galaxies 
become more dominant compared to the inner cluster region. For blue galaxies Agulli et al. finds $\alpha=-1.49^{+0.04}_{-0.03}$, which compares well 
to our $u$-band LF with $\alpha=-1.58\pm 0.12$. For the spectroscopic LF of A2151 from \citet{Agulli2017}, the faint-end slope was found to be 
$\alpha=-1.13\pm 0.02$ for the cluster-centric radius $0.0\leq r/r_{200}<1.4$ (using our adopted cosmology). This slope is flatter than our $r$-band 
slope of $\alpha=-1.38\pm 0.02$ for the full LF, and can be explained by a deficit of red dwarf galaxies in A2151 as reported by Agulli et al.

Comparing our results with the photometric LF of Coma from \citet{Beijersbergen2002}, we find good agreement with the faint-end slope for our full 
$r$-band LF ($\alpha=-1.38\pm 0.02$) and that of Beijersbergen et al. ($\alpha=-1.22^{+0.034}_{-0.036}$ for the cluster-centric annulus $0.0\leq r/r_{200}<1$, using our 
distance scale). For the $u$-band, Beijersbergen et al. finds $\alpha=-1.54^{+0.036}_{-0.030}$ for the annulus $0.0\leq r/r_{200}<0.5$, while for 
our full $u$-band LF we have $\alpha=-1.58\pm 0.12$. Since the $u$-band is expected to be more sensitive to recent star formation than the $r$-band 
\citep{Zhou2017}, we elect to compare our full $u$-band LF with the photometric UV LF from \citet{Cortese2003} for a composite sample of three 
low-redshift clusters (Virgo, Coma, and A1367). The faint-end slope of the Cortese et al. sample was determined to be $\alpha=-1.50\pm 0.10$, which is in 
good agreement with our $u$-band LF faint-end slope ($\alpha=-1.58\pm 0.12$).

\begin{figure}
\includegraphics[width=\linewidth]{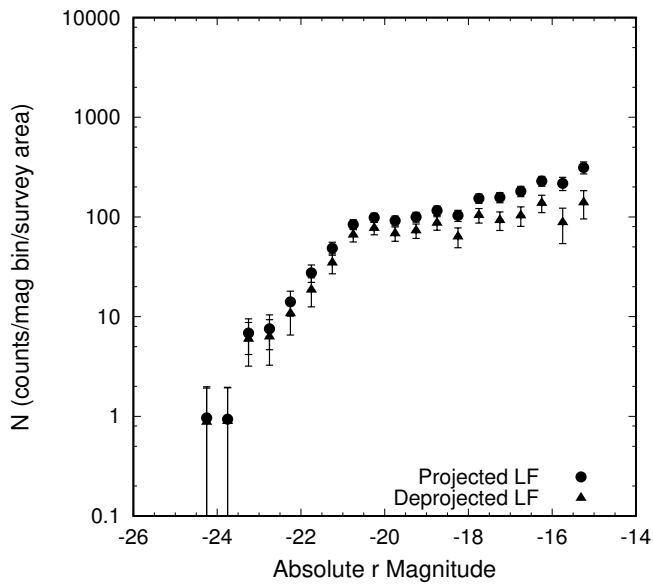}
\caption[Projected (original) and deprojected $r$-band LFs for the $0.0\leq r/r_{200} < 0.2$ region.]
{Projected and deprojected $r$-band LFs for the $0.0\leq r/r_{200} < 0.2$ region. Filled circles depict the 
projected LF, while the deprojected LF is represented by filled triangles.}
\label{0_to_20_rband_deprojected}
\end{figure}

\begin{figure}
\includegraphics[width=\linewidth]{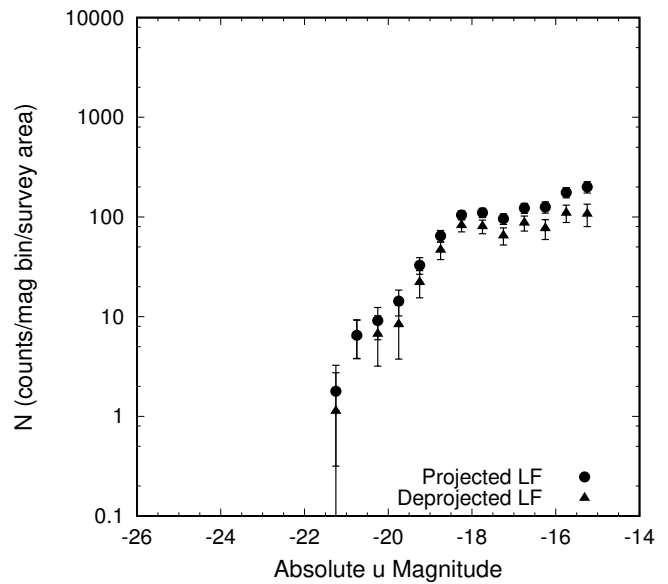}
\caption[Projected and deprojected $u$-band LFs for the $0.0\leq r/r_{200} < 0.2$ region.]{Projected 
and deprojected $u$-band LFs for the $0.0\leq r/r_{200} < 0.2$ region. Filled circles depict the 
projected LF, while filled triangles represent the deprojected LF.}
\label{0_to_20_uband_deprojected}
\end{figure}

\begin{figure}
\includegraphics[width=\linewidth]{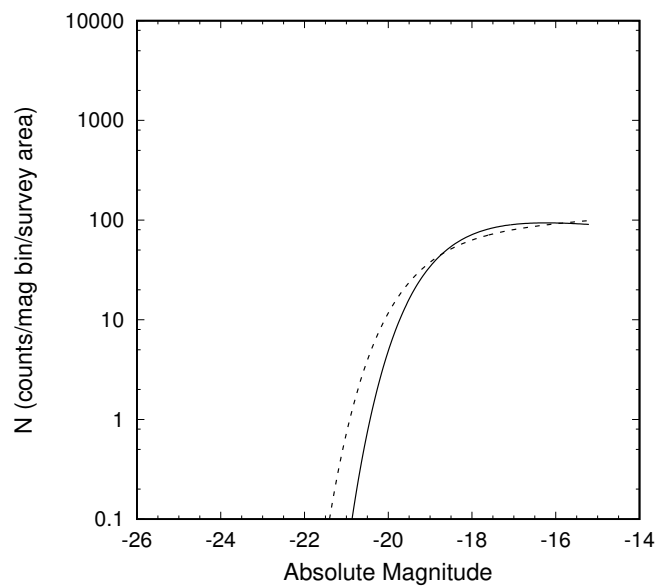}
\caption[Schechter function fits to the deprojected LFs in the $0.0\leq r/r_{200} < 0.2$ region.]{Schechter 
function fits to the deprojected LFs for the $0.0\leq r/r_{200} < 0.2$ region. The solid line is the fit to the 
$r$-band LF and the dashed line shows the fit to the $u$-band LF. The $u$-band data is shifted by 2.26 mag to the left for comparison.}
\label{deprojected_fits}
\end{figure}

An additional correction was also applied to study projection effects on the LFs. Assuming a cluster is 
spherically symmetric, the outer region of the cluster will be projected in front and behind the inner region of the 
cluster \citep{Beijersbergen2002,Barkhouse07}. This effect can be corrected for statistically by subtracting the 
contribution of the cluster outskirts from the inner cluster region. The resulting deprojected LFs are shown in 
Figs. \ref{0_to_20_rband_deprojected} and \ref{0_to_20_uband_deprojected}. In both cases, the resulting LFs are noticeably 
shallower at the faint-end than the projected LFs. In order to compare the two deprojected LFs, each LF was fit with a 
single Schechter function as neither LF shows an upturn of the faint-end slope. The two fits shown in Fig \ref{deprojected_fits} 
are consistent with each other, which indicates no enhanced star formation in the central region of the clusters.

\begin{table*}
\caption{Parameters derived from fitting double Schechter functions to the $u$-band LFs.}
\centering
\begin{threeparttable}
\centering
\begin{tabular}{|c|c|c|c|c|c|c|}
\hline
Radial Bin & $M_1^*$ & $\chi^{2}_{\nu}$ & $M_2^*$ & $\alpha_{2}$ & $\chi^{2}_{\nu}$ & No. of Clusters \\ \hline
$0.0 \leq r/r_{200} < 0.2$ & $-19.04 \pm 0.10$ & 1.82 & $-14.84 \pm 1.71$ & $-0.51 \pm 3.67$ & 1.63 & 15 \\ 
$0.2 \leq r/r_{200} < 0.4$ & $-18.98 \pm 0.08$ & 0.83 & $-15.16 \pm 0.13$ & $-1.58 \pm 0.81$ & 0.77 & 15 \\ 
$0.4 \leq r/r_{200} < 0.6$ & $-19.17 \pm 0.12$ & 1.17 & $-15.58 \pm 0.10$ & $-1.97 \pm 0.35$ & 1.41 & 15 \\
$0.6 \leq r/r_{200} < 1.0$ & $-19.13 \pm 0.09$ & 1.52 & $-16.17 \pm 0.10$ & $-1.65 \pm 0.10$ & 1.96 & 13 \\
$0.0 \leq r/r_{200} < 1.0$ & $-19.06 \pm 0.05$ & 1.30 & $-15.72 \pm 0.06$ & $-1.58 \pm 0.12$ & 1.10 & 13 \\ \hline
\end{tabular}
\end{threeparttable}
\label{tab:lf_fits_uband}
\end{table*}

\section{Dwarf-to-Giant Ratio and Blue Fraction}
\label{dwarf-to-giant_ratios_and_blue_fraction}

The DGR is used to search for evidence of star formation in a non-parametric way. For the $r$-band, giant galaxies 
are defined to have $M_{r} < -19.5$, and dwarfs have $-19.5 \leq M_{r} < -17.5$. Using an offset of 2.26 mag 
(Section \ref{sec:luminosity_functions}), giants in the $u$-band have $M_{u} < -17.24$ and dwarfs have 
$-17.24 \leq M_{u} < -15.24$. Since observations of Abell 2688 are not deep enough in the $u$-band to sample the 
dwarf regime, this cluster has been excluded from these measurements. The Abell 2688 background field, however, 
was used in the estimation of background counts for the lowest redshift clusters. To be consistent with previous 
measurements, the uncertainty in $N$ (galaxy count) is given by $\sqrt{N}$ (i.e. assuming Poisson statistics). 
When subtracting galaxy background counts from the cluster galaxy counts, uncertainties are added in quadrature.

\begin{figure}
\includegraphics[width=\linewidth]{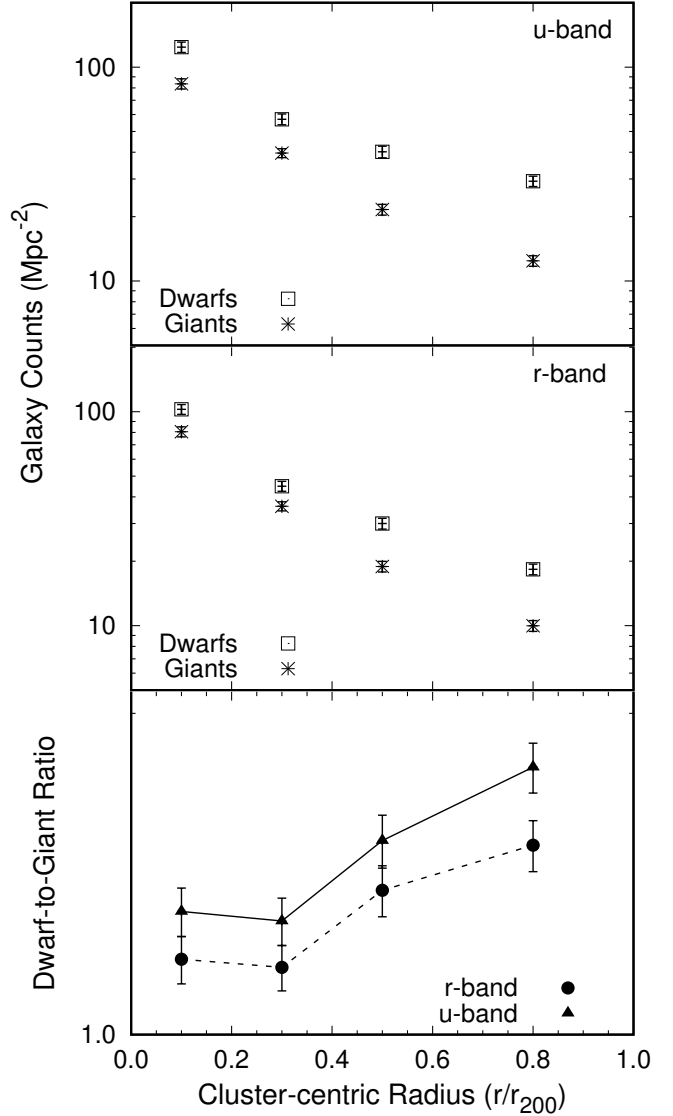}
\caption{Bottom: The $r$-band (filled circles with dashed line) and $u$-band (filled triangles with solid line) DGR for four radial bins. 
Middle: The $r$-band radial density profile for giant (asterisks) and dwarf (open squares) galaxies. Top: Radial density profile of 
giant (asterisks) and dwarf (open squares) galaxies measured in the $u$-band.}
\label{fig:dgr}
\end{figure}

The DGR as a function of $r/r_{200}$ is shown in Fig.~\ref{fig:dgr} (bottom panel). As expected from the LFs, the $u$-band DGR 
is marginally larger than the $r$-band in the inner cluster region (16 per cent difference), with both increasing 
with cluster-centric radius. In the outer region, the difference between the $u$- and $r$-band DGR increases 
to 28 per cent, consistent with a slight enhancement of star-forming dwarf galaxies. Both $u$- and $r$-band DGRs 
display a similar trend with cluster-centric radius, being relatively constant in the inner-most region and increasing 
in the cluster outskirts. This may be a result of dwarf galaxy disruption due to
gravitational tidal interactions and the quenching of star formation due to ram-pressure effects since the decline in the DGR is observed in both $u$ and $r$ filters 
\citep{Safarzadeh2017}.

To determine what gives rise to the change in the DGR with cluster-centric radius, we plot separately in Fig.~\ref{fig:dgr} the 
radial density profile of dwarfs and giants for the $r$-band (middle panel) and $u$-band (top panel). For both panels, the 
radial density decrease in dwarfs with increasing cluster-centric radius is less than for the giant galaxies. Thus the increase 
in the DGR toward the cluster outskirts (bottom panel) is due to a larger decrease in the density of giant galaxies compared 
to dwarf systems.

\begin{figure}
\includegraphics[width=\linewidth]{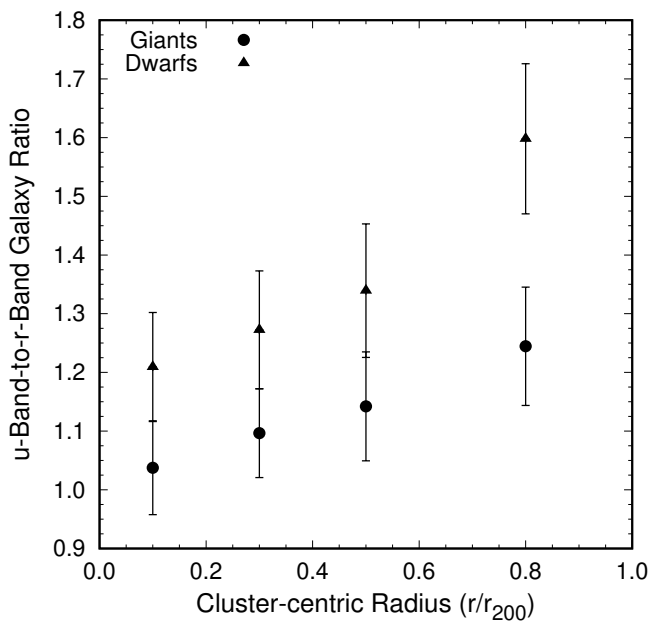}
\caption{Ratio of the number of $u$-band to $r$-band galaxies. Giants galaxies are depicted by filled circles, 
while dwarf systems are represented by filled triangles.}
\label{fig:u_to_r}
\end{figure}

One factor that may reduce the DGR in the $u$-band relative to the $r$-band is star formation in more luminous dwarf 
galaxies. A galaxy classified as a dwarf in the $r$-band may be considered a giant in the $u$-band. We explore this 
effect by plotting in Fig.~\ref{fig:u_to_r} the ratio of $u$-band to $r$-band detected galaxies. The inner region of 
the cluster sample contains a similar number of giants detected in both bands, while the outer region has 25 per 
cent more giants detected in the $u$-band. When comparing dwarfs, the $u$-band has 60 per cent more dwarfs than the 
$r$-band in the outer region.

\begin{figure*}
\includegraphics[width=\linewidth]{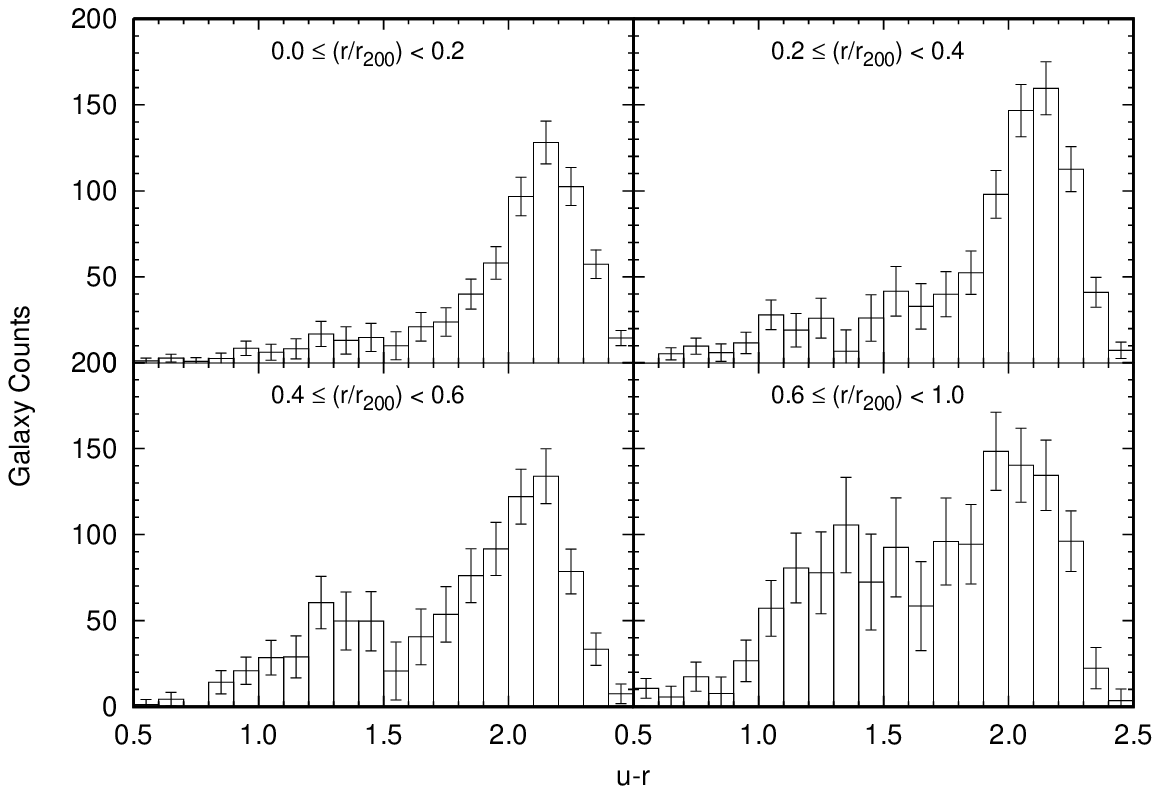}
\caption{Histogram colour distribution of dwarf galaxies within $-19.5 \leq M_r < -17$ in four radial bins.}
\label{color_hists}
\end{figure*}

To study changes in the dwarf galaxy population as a function of cluster-centric radius, histograms of $u-r$ colour are 
constructed for the four radial bins used previously. The error for each bin of the histogram is computed using 
$\sqrt{N_{ui} + N_{bi}}$, where $N_{ui}$ and $N_{bi}$ are the uncorrected cluster counts and the expected number of 
background counts in the $i$th bin, respectively. Dwarf galaxies are selected to be within $-19.5 \leq M_r < -17$. 
The results are shown in Fig.~\ref{color_hists}. 

\begin{figure}
\includegraphics[width=\linewidth]{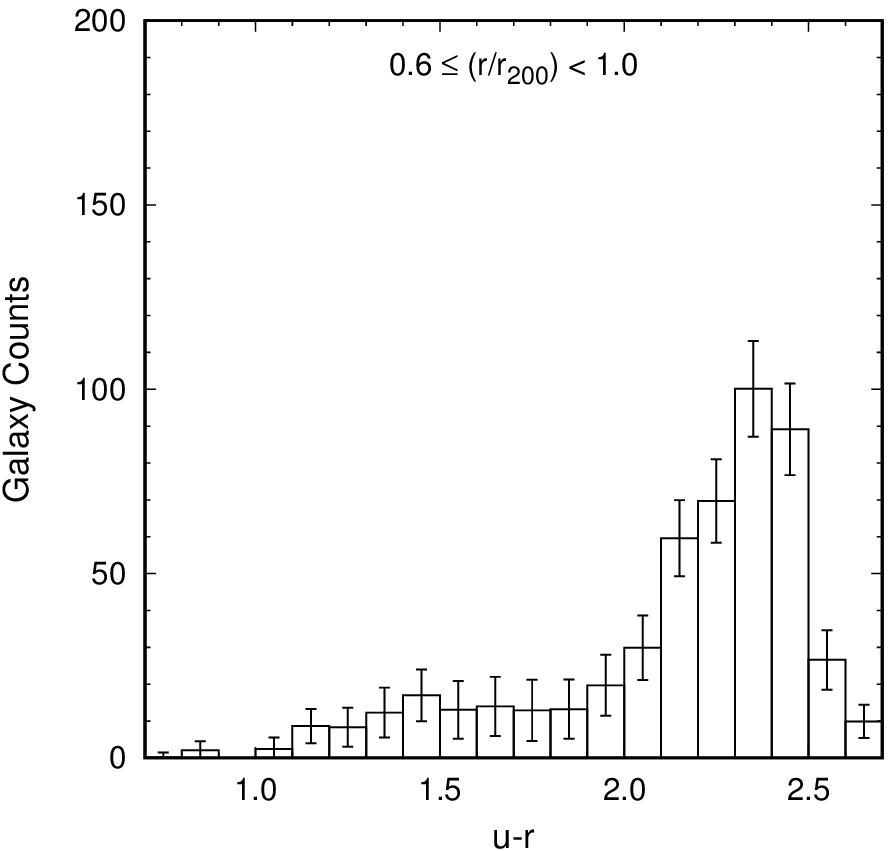}
\caption{Histogram colour distribution of giant galaxies within $-26 \leq M_r < -19.5$ for the $0.6 \leq r/r_{200} < 1.0$ region.}
\label{60_100_giants_color_hist}
\end{figure}

Comparing the four panels in Fig.~\ref{color_hists}, the relative size of the blue galaxy population grows relative to the red population 
with increasing cluster-centric radius. This coincides with the relative increase of the $u$-band relative to the $r$-band 
DGR. The colour of giant galaxies ($-26 \leq M_r < -19.5$) for the  $0.6 \leq r/r_{200} < 1.0$ region does not contain 
a large population of blue galaxies (Fig.~\ref{60_100_giants_color_hist}). 

Galaxies are considered to be part of the blue population if they are $>3\sigma$ blueward of the red-sequence. The 
blue fraction is given by $f_{b}=N_{b}/(N_{b}+N_{r})$, where $N_b$ is the number of background-corrected blue galaxies, 
and $N_r$ is the number of background-corrected red galaxies. The uncertainty in $N_b$ is determined by 
$\sigma_{N_{b}}=\sqrt{\Sigma \sigma_i^{2}}$, where the sum is over all bins blueward of the colour cut. The uncertainty 
in the number of red galaxies is calculated in a similar fashion. 

\begin{figure}
\includegraphics[width=\linewidth]{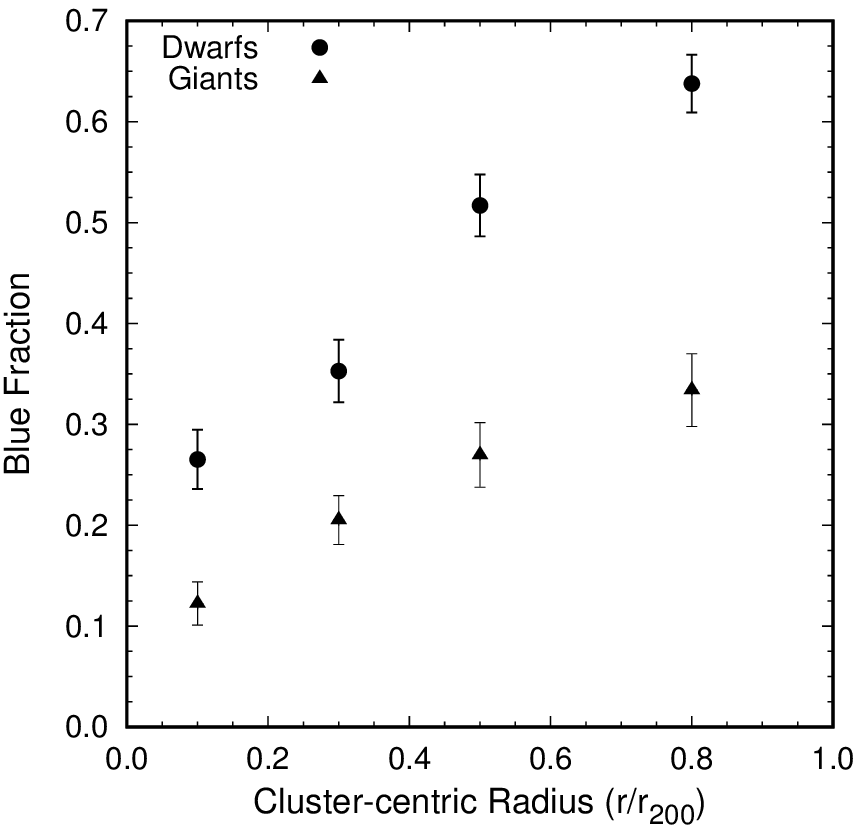}
\caption{Blue fraction versus cluster-centric radius for dwarfs (filled circles) and giants (filled triangles).}
\label{blue_fraction}
\end{figure}

The resulting dwarf and giant galaxy blue fractions for the four radial bins are depicted in Fig.~\ref{blue_fraction}. 
The blue fraction is greater for the dwarf population compared to giant galaxies by a factor of $\sim 2$ over all measured 
cluster-centric distance (significant at $\sim 5\sigma$ level). Both dwarf and giant galaxies undergo an increase in $f_{b}$ with increasing 
cluster-centric radius.

\section{Morphology}
\label{morphology}

Galaxies within $\pm 3\sigma$ of the red-sequence were characterised according to their $r$-band central concentration. The 
central concentration parameter ($C$) is a ratio of the flux measured using an inner and outer radius of a galaxy. 
It is determined by the intensity weighted second-order moments and defined as 
\citep{Abraham1994,Abraham1996} 
\begin{equation}
C = \frac{\sum_i \sum_{j\in E(\alpha)} I_{ij}}{\sum_i \sum_{j\in E(1)} I_{ij}} , 
\end{equation}
where ${I_{ij} }$ is the intensity of a pixel in position ($i$, $j$), $E$($\alpha$) is the inner normalised elliptical 
radius, and $E(r=1)$ is the outer elliptical radius normalised to 1. The inner radius isolates the flux 
within the cores of galaxies, and $\alpha$=0.3 has been found empirically to produce the best results 
\citep{Abraham1994}. 

\begin{figure}
\includegraphics[width=\linewidth]{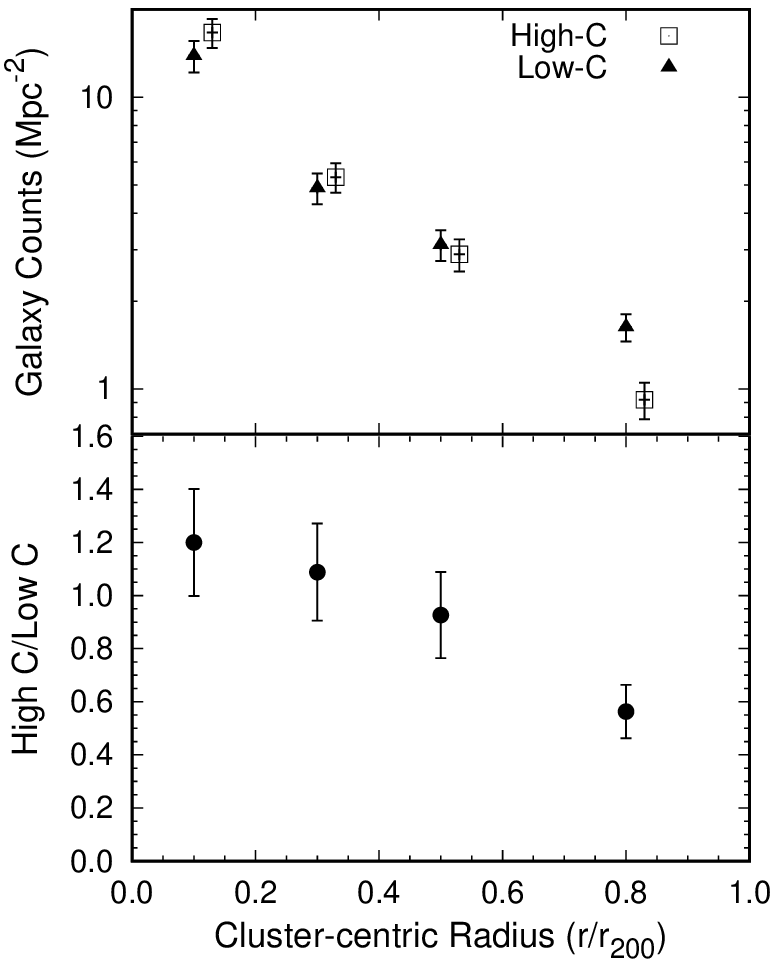}
\caption{Bottom: Ratio of high-$C$ to low-$C$ dwarf galaxies as a function of cluster-centric radius ($r/r_{200}$). 
Top: The radial density distribution of high-$C$ (open squares) and low-$C$ (solid triangles) dwarf galaxies. 
The high-$C$ $r/r_{200}$ values have been offset slightly to better show the difference in the data points.}   
\label{HighCLowC}
\end{figure}

\begin{figure}
\includegraphics[width=\linewidth]{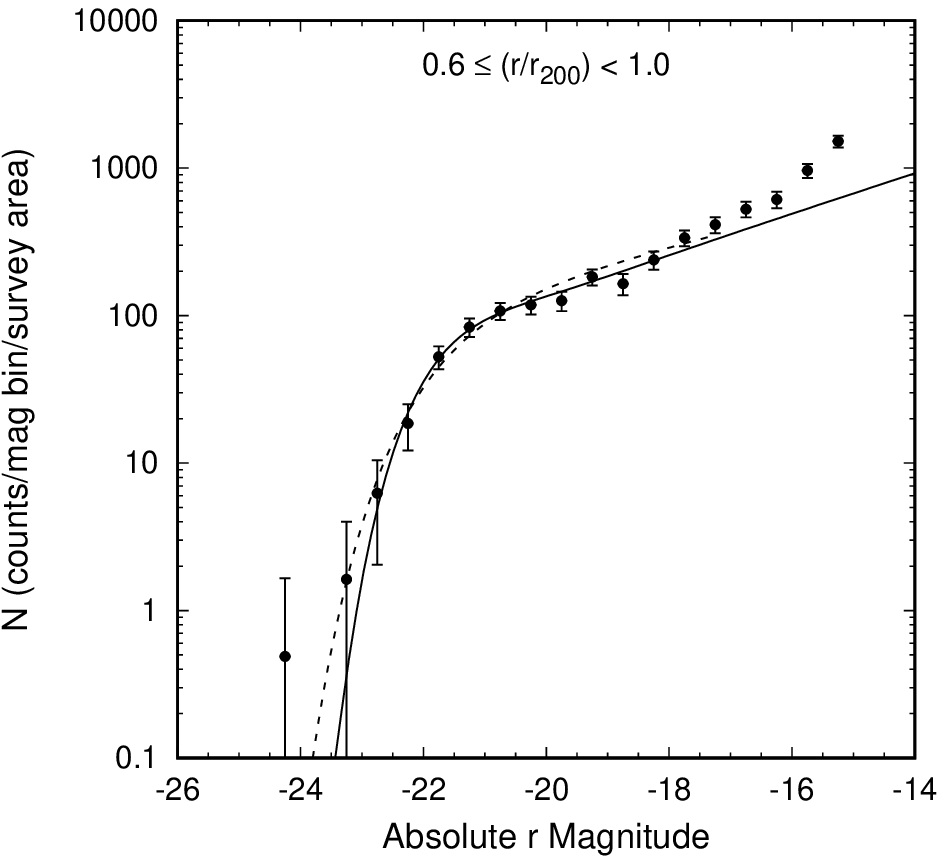}
\caption[The $r$-band LF for the $0.6 \leq r/r_{200} < 1.0$ region of the CFHT clusters, 
shown with the field LF from Blanton et al. (2005; solid line).]{The composite $r$-band LF 
for the $0.6 \leq r/r_{200} < 1.0$ region of the CFHT clusters, shown with the field LF from Blanton et al. 
(2005; solid line) and Montero-Dorta \& Prada (2009; dashed line). The field LFs have been converted to our adopted 
cosmology and scaled to match the bright-end of the CFHT LF.}
\label{fig:blanton_r_lf}
\end{figure}

\begin{figure}
\includegraphics[width=\linewidth]{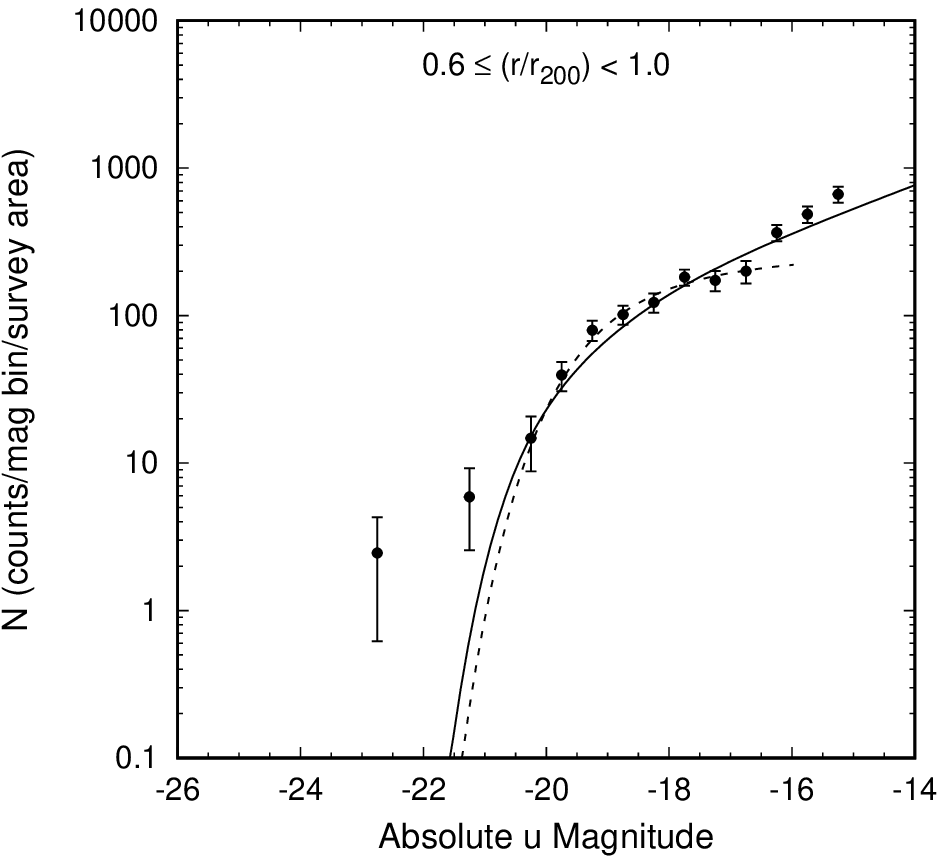}
\caption[The composite $u$-band LF for the $0.6 \leq r/r_{200} < 1.0$ region of the CFHT clusters, 
shown with the field LF from Blanton et al. (2005; solid line).]{The composite $u$-band LF for the 
$0.6 \leq r/r_{200} < 1.0$ region of the CFHT clusters, shown with the field LF from Blanton et al. 
(2005; solid line) and Montero-Dorta \& Prada (2009; dashed line). The field LF has been converted 
to our adopted cosmology and scaled to match the bright-end of the CFHT LF.}
\label{fig:blanton_u_lf}
\end{figure}

Measurements of the FWHM for stars in each cluster image were used to define a minimum isophotal area 
for classification. Galaxies having a diameter less than $3\times$FWHM are considered too small for accurate 
morphological classification. 

The central concentration values for dwarf galaxies ($-19.5 \leq M_r < -17$) from all 15 clusters were measured. 
The dwarf galaxies were separated into two, approximately equal samples: low central concentration ($C<0.27$) 
and high central concentration ($C\geq 0.27$).

Fig. \ref{HighCLowC} shows the ratio of high-$C$ versus low-$C$ dwarf galaxies as a function of cluster-centric 
radius (bottom panel), and the radial density distribution of low-$C$ and high-$C$ dwarf galaxies (top panel). 
Uncertainties are calculated assuming Poisson statistics, and are added in quadrature. The central concentration ratio 
for the inner-most radial bin is 2.1 times greater than the outer-most annuli, significant at the 2.8$\sigma$ level. The 
radial density profiles indicate that the density of high-$C$ dwarf galaxies decreases more rapidly than the low-$C$ 
galaxies with increasing cluster-centric radius.

For this study we associate high-$C$ dwarf galaxies with nucleated dwarfs and low-$C$ dwarfs with non-nucleated 
galaxies \citep[e.g.][]{vandenbergh86}. Gravitational tidal perturbations in cluster centres are expected 
to have the greatest effect on non-nucleated, loosely bound low-$C$ galaxies compared to more centrally concentrated, 
high-$C$ nucleated systems \citep{Conselice2001,Eigenthaler2010}. Thus we expect that the ratio of high-$C$ to 
low-$C$ dwarf galaxies would increase towards the cluster centre, as evident in Fig. \ref{HighCLowC}. 

\citet{Lisker2007} and \citet{Ordenes2018} mapped out the distribution of nucleated and non-nucleated 
dwarf galaxies in the Virgo and Fornax clusters, respectively. These studies showed that the ratio of nucleated to 
non-nucleated dwarfs increases toward the cluster centre. These results are consistent with the change in the ratio 
of high-$C$ to low-$C$ dwarf galaxies depicted in Fig. \ref{HighCLowC} for our composite sample of 15 galaxy clusters.

\section{Discussion and Conclusions}
\label{discussion}

The faint-end of LFs measured for galaxies in the low-density field environment differs from LFs 
measured from the outskirts of the cluster region ($0.6 \leq r/r_{200} < 1.0$). Comparison between the LF from the 
outermost region of our composite cluster sample and the SDSS field LF from \citet{Blanton05} is shown in 
Figs.~\ref{fig:blanton_r_lf} ($r$-band) and \ref{fig:blanton_u_lf} ($u$-band). The field LF is based on SDSS DR2 and 
has been normalised to match the bright-end of the CFHT LF. We use the ``raw'' field LF from Blanton et al. since 
it most-closely matches how our galaxies are selected, in contrast to the ``corrected'' or ``total'' SDSS 
field LF. In Figs.~\ref{fig:blanton_r_lf} and \ref{fig:blanton_u_lf} we also compare our cluster LF to the field 
LF from \citet{Montero09}. Our $r$-band cluster LF is steeper ($\alpha=-1.63\pm 0.02$) than the SDSS field LF from Blanton et al. 
($\alpha=-1.34\pm 0.01$). For the $u$-band, our cluster LF faint-end slope ($\alpha=-1.65\pm 0.10$) is marginally steeper than the 
Blanton et al. SDSS field LF ($-1.39\pm 0.02$) at the $2.5\sigma$ level. 

The similarity of the faint-end slope of the cluster $u$-band LF compared to the SDSS field LF indicates that there is 
no significant enhancement of star formation as dwarf galaxies fall into the cluster environment from the field. 
The shallower $r$-band faint-end slope of the field LF compared to the cluster $r$-band LF may imply 
that star formation in dwarfs is being quenched, ultimately transforming blue, star-forming dwarf galaxies into 
red, passive systems. 

The change in the $r$-band faint-end slope with cluster-centric radius may imply that the survival rate of dwarf galaxies 
is highest in the outskirts of clusters compared to the inner cluster region. This is 
expected as dwarf galaxies are likely to be destroyed by gravitational tidal effects and ram pressure stripping in 
denser environments \citep{Martel2012,Safarzadeh2017,Zinger2018}. Consistent with these results, \citet{Thompson93} 
found that less centrally concentrated dwarf galaxies (dSph) are absent in the centre of the Coma cluster. Our 
measurement of an increasing ratio of high-$C$ to low-$C$ dwarf galaxies with decreasing cluster-centric radius also 
supports this view.

The $r$-band DGR versus cluster-centric radius for the CFHT clusters increases for  $r/r_{200}\gtrsim 0.4$. The DGR 
selected from a sample of 57 low-redshift Abell clusters \citep{Barkhouse09} shows a steady 
increase from the inner to the outer region. Differences in the DGR may indicate systematic differences between 
estimations of $r_{200}$ between \citet{Barkhouse09} and our sample. The richness method used to estimate 
$r_{200}$ was different ($\lambda$ vs $B_{gc}$), as was the sample of clusters used to generate the 
relationship between richness and $r_{200}$. Both cases, however, indicate a suppression of dwarf galaxies in 
the inner region of low-redshift clusters. Even if there are systematic differences between the two samples, 
comparisons within the CFHT data is valid as $r_{200}$ is calculated in the same way for all CFHT 
clusters. Additionally, Barkhouse et al. used the $B$-band for their blue filter. The two filters ($u$ and $B$) 
may sample different stellar populations and thus the $B$-band may indicate star formation on a 
different timescale \citep{Larson1978}. 

The measured blue fractions of the two studies also differ. While our results show a steady increase in the dwarf 
blue fraction with increasing radius, Barkhouse et al. found an increase in the dwarf blue fraction at small radii, 
but a levelling off for $r/r_{200}<0.4$. If star formation in infalling dwarf galaxies is being quenched 
in the outskirts of clusters, it would not show up as quickly in the $B$-band.

The relative location of star formation in a cluster can only be determined if the quenching of star formation occurs 
over a short time period. Galaxy starvation is effective on timescales of $\approx10^9$ years, where as ram pressure 
could remove the gas within a disk in $10^8$ years \citep{Quilis2000}, though some simulations show it could take much 
longer \citep{Tonnesen2007}. The change in the blue fraction and LFs with respect to cluster-centric radii implies 
quenching timescales $< 10^9$ years.

\label{summary}
\section*{Acknowledgements} 
We thank the anonymous referee for comments that improved this manuscript. We thank Haylee Archer and Madeline Shaft 
for assistance with the visual inspection of detected objects. CMR and WAB thank the University of North Dakota for 
financial support through a ND EPSCoR Doctoral Dissertation Assistantship and a Faculty Research Seed Money award. 
MRS and WAB acknowledge financial support from the Theodore Dunham, Jr. Grant of the Fund for Astrophysical Research. 
This research has made use of the VizieR catalogue access tool, CD'S, Strasbourg, France. This research made use of 
the K-corrections Calculator service available at \url{http://kcor.sai.msu.ru/}, and GNU Parallel \citep{Tange2018}. 
Based on observations obtained with MegaPrime/MegaCam, a joint project of CFHT and CEA/DAPNIA, at the Canada-France-Hawaii 
Telescope (CFHT) which is operated by the National Research Council (NRC) of Canada, the Institut National des Sciences 
de l'Univers of the Centre National de la Recherche Scientifique of France, and the University of Hawaii.



\bsp	
\label{lastpage}
\end{document}